\documentclass[aps,prd,floatfixy,superscriptaddress,showpacs,showkeys]{revtex4}
\usepackage{graphicx}
\usepackage{booktabs}
\usepackage{amssymb,bm,mathrsfs,bbm,amscd}
\usepackage[tbtags]{amsmath}
\usepackage{lastpage}
\newcommand{\eq}{\begin{eqnarray}}
\newcommand{\en}{\end{eqnarray}}

\newcommand{\SL}{\rlap{\slash }}

\begin{document}

\title{Deuteron form factors in a phenomenological approach}

\author{Cuiying Liang and Yubing Dong}
\affiliation{
Institute of High Energy Physics, Chinese Academy of Sciences,
Beijing 100049, China}
\affiliation{
Theoretical Physics Center for Science Facilities (TPCSF), CAS,
Beijing 100049, China}

\date{\today}

\begin{abstract}
The electromagnetic form factors of the deuteron, particularly the quadrupole
form factor,  are  studied with a help of a phenomenological Lagrangian approach where
the vertex of the deuteron-proton-neutron with $D$-state contribution is
explicitly taken into account. The result shows the importance of this
contribution to the quadrupole form factor in the approach.
\end{abstract}



\maketitle


\section{Introduction}
The study of electromagnetic form factors of nucleon and light nuclei, like
deuteron and He-3, are crucial for the understanding of the nucleon
structures. Deuteron, as the most simplest nuclei, has
been a subject of many years (for some recent
reviews\cite{Garcon, Gilman, Gross, Sick}). Since it is a weekly bound state of
the proton and neutron, the study of the deuteron can shed light on the study
of the nucleon as well as on the nuclear effects. 
Moreover, as a spin-1 particle,
the deuteron structures are different from the spin-1/2 nucleon and He-3, 
and from
the spinless pion meson. There are many discussions on the deuteron structures,
like its wave functions, binding energy, the electromagnetic form factors, and
the parton distributions, in the literature. Those works are
usually based on the phenomenological potential models with quark, meson,
and nucleon degrees of freedom and based on some effective field theories etc.
\cite{Garcon, Gilman, Gross, Sick, Arnold, Mathiot, Wiringa, Arhen, Gari,
Karm, Kaplan, Ivanov}. The realistic deuteron wave functions, with the help
of meson exchange potential model, have been explicitly given by
Refs. \cite {Mau, Gross1, Carbonell}.

In our previous works \cite{Dong0, Dong1}, a phenomenological Lagrangian
approach is applied for the electromagnetic form factors of the deuteron,
where it is regarded as a loosely bound state of a proton and a neutron, and
the two constituents are in relative $S$-wave for simplicity. The coupling
of the deuteron to its two composite particles is determined by the known
compositeness condition from Weinberg \cite {Weinberg}, Salam \cite {Salam}
and others \cite {Hay, Efi}. Our phenomenological effective Lagrangian approach
has been proven to be successful in the study the weekly bound state problems,
like the new resonances of $X(3872)$, and $\Lambda_c^+(2940)$, the EM form 
factors
of pion as well as some other observables  \cite{Dong2, Dong3}.

It should be stressed that since only one-body $S-$ wave operator contribution
is considered in our previous study \cite{Dong0}, the estimated quadrupole 
moment of the deuteron is negligibly small when compared to data.  
According the non-relativistic potential model calculation \cite {Mau}, one 
sees that the deuteron quadrupole moment is very sensitive to the $D-$ wave 
component of  the deuteron. Therefore, $S-$ state contribution is not 
sufficient. In order to avoid the discrepancy, several two-body arbitrary 
and phenomenological Lagrangians were introduced, by hand, to compensate the 
discrepancy \cite{Dong0}.

The purpose of this work is to re-study the deuteron electromagnetic
form form factors with the phenomenological approach. Here both the $S-$
and $D-$ state contributions to the vertex of the deuteron-proton-neutron are
simultaneously taken into. It is expected that by considering the $D-$
state contribution in the vertex, the estimated deuteron quadrupole could
be sizeably improved. This paper is organized as follows. Section 2 briefly 
shows our theoretical framework, particularly the $D-$ state contribution 
to the vertex. Numerical results and some discussions are given in section 3.

\section{Framework of the approach}
\begin{figure}
\begin{center}
\includegraphics[width=5cm,height=3cm]{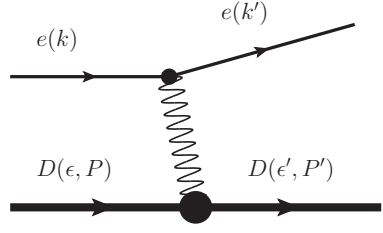}
\caption{Electron-deuteron scattering}
\end{center}
\end{figure}
Deuteron,  as a spin-1 particle, has three independent form factors. The
matrix element for electron-deuteron (ED) elastic scattering, as shown in
Fig. 1, can be written as
\eq
{\cal M}=\frac{e^2}{Q^2} \bar u_e(k^\prime) \gamma_{\mu}
u_e(k) {\cal J}_{\mu}^D(P,P^\prime),
\en
under the one-photon exchange approximation. In eq. (1) $k$ and $k^\prime$
are the four--momenta of initial and final electrons and
${\cal J}_{\mu}^D(P,P^\prime)$ stand for the deuteron EM current. Its
general form is
\eq\label{D_current}
&&{\cal J}_{\mu}^D(P,P^\prime)=\\ \nonumber
&&-\biggl( G_1(Q^2)\epsilon^{\prime *}\cdot\epsilon-\frac{G_3(Q^2)}{2M_d^2}
\epsilon\cdot q\epsilon^{\prime *}\cdot q \biggr) (P + P^\prime)_{\mu}
\\ \nonumber
&&-G_2(Q^2) \biggl( \epsilon_{\mu}\epsilon^{\prime *}\cdot q
- \epsilon^{\prime *}_{\mu} \epsilon\cdot q \biggr)~,
\en
where $M_d$ is the deuteron mass, $\epsilon$($\epsilon^\prime$) and
$P(P^\prime)$ are polarization and four--momentum of the initial (final)
deuteron, and $Q^2=-q^2$ is momentum transfer square with $q=P^\prime - P$.
The three EM form factors $G_{1,2,3}$ of the deuteron are related to the
charge $G_C$, magnetic $G_M$, and quadrupole $G_Q$ form factors by
\eq
G_C&=& G_1+\frac23\tau G_Q\,, \hspace*{.25cm}
G_M \ = \ G_2 \,,            \hspace*{.25cm}\\ \nonumber
G_Q&=&G_1-G_2+(1+\tau)G_3,   \hspace*{.25cm}
\en
with the factor of $\tau=Q^2/4M_d^2$. These three form factors are
normalized at zero recoil ($Q^2=0$) as
\eq
G_C(0)&=&1\,, \ \ \
G_Q(0)=M_d^2{\cal Q}_d=25.83\,, \\ \nonumber
G_M(0)&=&\frac{M_d}{M_N}\mu_d=1.714 \, ,
\en
where $M_N$ is the nucleon mass, ${\cal Q}_d$ and $\mu_d$ are the quadrupole
and magnetic moments of the deuteron.

The unpolarized differential cross section for the $eD$ elastic
scattering can be expressed by the two structure functions, $A(Q^2)$ and
$B(Q^2)$, as
\begin{equation}
{{d\sigma} \over {d\Omega}}=\sigma_M \left[ A(Q^2)
+B(Q^2) \tan^2{\left(\frac{\theta}{2}\right)} \right ] ,
\label{eq:eq10}
\end{equation}
where $\sigma_M=\alpha^2 E^\prime \cos^2(\theta/2)/[4 E^3 \sin^4(\theta/2)]$
is the Mott cross section for point-like particle, $E$ and $E^\prime$
are the incident and final electron energies, $\theta$ is the electron
scattering angle, $Q^2=4 E E^\prime \sin^2(\theta/2)$,
and $\alpha=e^2/4\pi=1/137$ is the fine-structure constant.
The two form factors $A(Q^2)$ and $B(Q^2)$ are related to the three EM form
factors of the deuteron as
\eq
&&A(Q^2)=G^2_C(Q^2)+\frac89\tau^2 G^2_Q(Q^2)+
\frac23\tau G^2_M(Q^2) \nonumber \\
&&B(Q^2)=\frac43\tau (1+\tau) G^2_M(Q^2) .
\en
Clearly, the three form factors $G_{C,M,Q}$ cannot be simply determined by
measuring the unpolarized elastic $eD$ differential cross section.
To uniquely determine the three form factors of the deuteron one additional
polarization variable is necessary. For example, one may take the polarization
of $T_{20}$ \cite{Garcon}.

Take an assumption that the deuteron as a hadronic molecule--a weakly bound
state of the proton and neutron, one may simply write a
phenomenological effective Lagrangian of the deuteron and its two
constituents of the proton and neutron as
\eq
{\cal L}_D(x)=g_DD_{\mu}(x)\int
dy\Phi_D(y^2)\bar{p}(x+y/2) \Gamma^{\mu}_{D}C\bar{n}^{T}(x-y/2)
+H.c.,
\en
where $D_{\mu}$ is the deuteron field, $C\bar{n}^T(x)=n^c(x)$, and
$C=i\gamma^2\gamma^0$ denotes the matrix of charge conjugation, and x is the
centre-of-mass
(C. M.) coordinate. In the above equation $\Gamma_D^{\mu}$ is the vertex
for the deuteron-proton-neutron and the correlation function $\Phi_D(y^2)$
characterizes the finite size of the deuteron as a $pn$ bound state.
The correlation function $\Phi_D(y^2)$ depends on the relative
Jacobi coordinate $y$.

If only the $S-$ wave contribution is considered, the simplest form of the
vertex is $\Gamma_{D}^{\mu}\sim\gamma^{\mu}$ which has been employed before
\cite{Dong0}. When both the $S-$ and $D-$ states contributions are considered,
then the vertex becomes more complicated. According to the work of
Blankenbecler, Gloderber, and Halpern \cite {Goldberger} the vertex of the
deuteron-proton-neutron is
\eq
\Gamma_{D}^{\mu}=\Gamma_{D}^{1,\mu}+\Gamma_{D}^{2,\mu}
\en
where the first and second terms stands for the contribution from $S-$ and
$D-$ states, respectively. They are
\eq
\Gamma_{D}^{1,\mu}=\frac{1}{2\sqrt{2}}\Big (1+\frac{\SL{P}}{M_d}\Big )
\gamma^{\mu}
\en
and
\eq
\Gamma_{D}^{2,\mu}=\frac{\rho}{16}\Big (1+\frac{\SL{P}}{M_d}\Big )
\Big (\gamma^{\mu}-\frac{3}{k^2}\SL{k}\gamma^{\mu}\SL{k}\Big )
\en
with $\rho$ being a measure of the $D-$state admixture, $k$ is the relative
momentum between the proton and neutron, and $k^2=M_N\delta$ with $\delta$
being the binding energy of the deuteron as shown in Fig. 2. Here, it should
be mentioned that in the rest frame of the deuteron, the non-relativistic
reduction gives
\eq
\epsilon_{i} \Gamma_D^{1,i}C=-\frac{i}{\sqrt{2}}\vec{\sigma}
\cdot\vec{\epsilon}\sigma_2=
\left (\begin{matrix}
\epsilon_{-1} &-\frac{1}{\sqrt{2}}\epsilon_z \\
-\frac{1}{\sqrt{2}}\epsilon_z &\epsilon_{+1}
\end{matrix} \right).
\en
It means a combination of two spin-1/2 states, proton and neutron, forms a
spin triplet state. Similarly, in the non-relativistic limit,
$\Big (\gamma^{\mu}-\frac{3}{k^2}\SL{k}\gamma^{\mu}\SL{k}\Big )$
means the proton and neutron couple to spin triplet state and this spin
triplet state re-couples $Y_{2m_l}(\hat{k})$ to form a state with the same
quantum numbers of the deuteron.

\vspace{0.5cm}
\begin{figure}\begin{center}
\includegraphics[width=7.5cm, height=2cm]{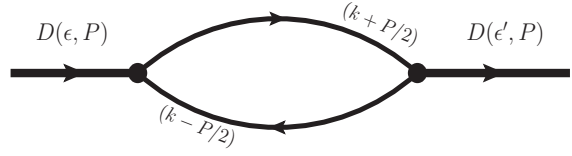}
\caption{The mass operator of the deuteron}
\end{center}\end{figure}
\vspace{0.5cm}
The coupling of the deuteron to its two constitutes, $g_D$ in eq. (7)
is determined by the known compositeness condition $Z=0$ proposed by
Weinberg, Salam and others \cite {Weinberg,Salam,Hay,Efi}. This condition
implies that the probability to find a proton and neutron system inside the
deuteron is unity. Thus, the coupling of $g_D$ is determined according to
$Z_D=1-\Sigma_D'(M_D^2)=0$, with
\eq
\Sigma_D'(M_D^2)=g_D^2\Sigma_{D\perp}'(M_D^2)
\en
being the derivative of the transverse part of the mass operator (see Fig. 2).
Usually, the mass operator splits into the transverse and longitudinal parts
of $\Sigma^{\alpha\beta}_D(k)=g^{\alpha\beta}_{\perp}\Sigma_{D\perp}(k^2)+
\frac{k^{\alpha}k^{\beta}}{k^2}\Sigma_{D\parallel}(k^2),$
with $g_{\perp}^{\alpha\beta}=g^{\alpha\beta}-k^{\alpha}k^{\beta}/k^2$
and $g_{\perp}^{\alpha\beta}k_{\alpha}=0$.  We see that the
coupling of the deuteron to its constituents of the proton and neutron,
$g_D$, is well determined by the compositeness condition.

A basic requirement for the choice of an explicit form of this correlation
function is that its Fourier transform vanishes sufficiently fast in the
ultraviolet region of Euclidean space to render the Feynman diagrams
ultraviolet finite. Usually a Gaussian-type function is selected as
the correlation for simplicity. One may choose ${\tilde \Phi}_D(k^2)
=exp(-k_E^2/\Lambda^2)$ for the Fourier transform of the correlation
function, where $k_E$ is the Euclidean Jacobi relative momentum and $\Lambda$
stands for the free size-parameter which represents the distribution of the
constituents in the deuteron.

Here the analytical expression for the coupling is
\eq
\frac{1}{g^2_D}=\Sigma_{D\perp,1}'+\rho\Sigma_{D\perp,2}',
\en
where $\Sigma_{D\perp,1}'$ and $\Sigma_{D\perp,2}'$ stand for the derivatives
of the transverse parts of the mass operator from the contributions of the
$S-$ and $D-$ states, respectively. The explicit expressions are
\eq
\Sigma_{D\perp,1}'&=&\frac{1}{32\pi^2}
\int \frac{d\alpha d\beta}{Z_0^3}\times \Bigg \{
\frac{A(\alpha,\beta)}{Z_0}\Big [1+\frac{\Lambda_S^2}{4M_d^2Z_0}\Big ]
\nonumber \\
&&+\frac{B(\alpha,\beta)}{2}\times \Big [
\mu_d^2\Big (1+\frac{A(\alpha,\beta)}{Z_0^2}\big (1+\frac{\Lambda_s^2}{
4M_d^2Z_0}\big )\Big )
+\frac{3\Lambda_S^2}{2M_d^2Z_0^2}-\frac{1}{4Z_0}\Big ]\Bigg \}\nonumber \\
&&\times exp\Big [-2(\alpha+\beta)\mu_N^2
+\frac{A(\alpha,\beta)}{2Z_0}\mu_d^2\Big ]
\en
where $\mu_{N,d}=M^2_{N,d}/\Lambda_S^2$ and
\eq
A(\alpha,\beta)&=&(1+2\alpha)(1+2\beta)\\ \nonumber
B(\alpha,\beta)&=&\alpha+\beta+4\alpha\beta\\ \nonumber
Z_0&=&1+\alpha+\beta,
\en
and
\eq
\Sigma'_{D\perp,2}&=&\int\frac{d\alpha d\beta}{16\sqrt{2}\pi^2Z_1^3}
\times \Bigg \{
\frac{A'(\alpha,\beta)}{Z_1}\Big [1+\frac{3\Lambda_S^2}{8\delta M_DZ_1}\Big ]
\nonumber \\
&&+\frac{B'(\alpha,\beta)}{2}\Big [
\mu_d^2\Big (1+\frac{A'(\alpha,\beta)}{Z_1^2}
\big (1+\frac{3\Lambda_S^2}{8\delta M_dZ_1}\big ) \Big )
+\frac{1}{2Z_1^2}
\Big (1-\frac{15M_d}{8\delta}+\frac{9\Lambda_S^2}{2\delta M_d}
\Big ) \Big ] \Bigg \}\nonumber \\
&&\times exp\Big [-2(\alpha+\beta)\mu_N^2
+\frac{A'(\alpha,\beta)}{2Z_1}\mu_d^2\Big ]
\en
with
\eq
&&A'(\alpha,\beta)=\big (\frac{1+a_{SD}}{2}+2\alpha\big )
\big (\frac{1+a_{SD}}{2}+2\beta\big )
\nonumber \\
&&B'(\alpha,\beta)=\frac{1+a_{SD}}{2}(\alpha+\beta)+4\alpha\beta \nonumber \\
&&Z_0=\frac{1+a_{SD}}{2}+\alpha+\beta,
\en
and $a_{SD}=\Lambda_S^2/\Lambda_D^2$. Here we simply ignore the $\rho^2$-
dependent term since $\rho$ is expected to be small, and we consider that
the correlation functions of the $S-$ and $D-$ states are not necessarily
the same, therefore we have totally three parameters $\Lambda_S$,
$\Lambda_D$ and $\rho$ in this calculation.
\begin{figure}\begin{center}
\centering
\includegraphics[width=7.5cm, height=3cm]{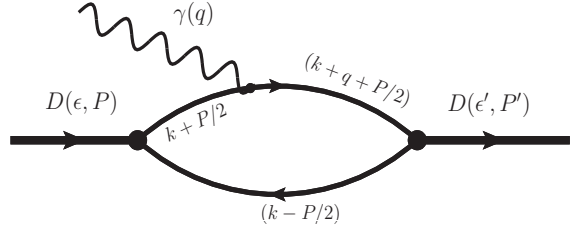}
\caption{Photon-deuteron interaction}
\end{center}\end{figure}
Then, we can calculate the matrix element of photon-deuteron interaction as
shown in Fig. 3. We have
\eq
{\cal M}^{\mu}&=&\sum_{(N=p,n)}~~\sum_{(i,j=1,2)}
\int \frac{d^4k}{(2\pi)^4i}g_D^2\epsilon^{'*}_{\alpha}\epsilon_{\beta}
\times Tr\Big [ \frac{\Gamma_{D}^{i,\alpha}(\SL{k}+\SL{q}+\SL{p}/2+M_N)}
{(k+q+p/2)^2-M^2_N}\nonumber \\
&&\cdot \frac{\Gamma_{\gamma N}^{\mu}
(\SL{k}+\SL{p}/2+M_N)}{(k+p/2)^2-M_N^2}\cdot
\frac{\Gamma_{D}^{j,\beta}(\SL{k}-\SL{p}/2-M_N)}
{(k-p/2)^2-M^2_N}\Big ]\nonumber \\
&&\times exp\Big [-k_E^2/\Lambda^2_j-(k+q/2)_E^2/\Lambda_i^2\Big ]
\en
where the photon-nucleon current of
\eq
\Gamma_{\gamma N}^{\mu}=F_{1,N}(Q^2)\gamma^{\mu}
+F_{2,N}(Q^2)\frac{i\sigma^{\mu\nu}}{2M_N}q_{\nu}
\en
is employed with $F_{1,N}$ and $F_{2,N}$ being the known nucleon Dirac and
Pauli form factors and $N=p,n$, stand for the proton and neutron,
respectively.

\section{Numerical results and discussions}

We calculate the matrix element of eq. (18) and consider the one-photon
exchange approximation for the photon-deuteron current as shown eq. (1).
Thus we can get the corresponding deuteron three form factors $G_{1,2,3}$
as well as the deuteron charge $G_c(Q^2)$, magnetic $G_M(Q^2)$ and 
quadrupole $G_Q(Q^2)$
form factors. There are some parameterizations for the nucleon
form factors of $F_{1,2}(Q^2)$ in the literature for the proton and neutron
by \cite{Drechsel,Kelly,Blunden}. In the present calculation, we employ the
ones of Blunden \cite{Blunden}. The three model-dependent parameters,
$\Lambda_S=0.10~GeV$, $\Lambda_D=0.08~GeV$ and $\rho=0.03$, are fixed by
fitting  to the experimental data. The obtained charge, magnetic and
quadrupole form factors are shown in Figs. (4-6).
\vspace{1.cm}
\begin{figure}\begin{center}
\centering
\includegraphics[width=7.5cm, height=6cm]{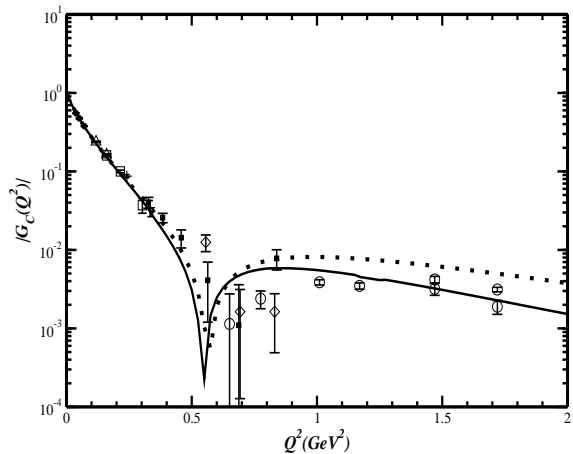}
\caption{Estimated deuteron charge form factor $G_c(Q^2)$.
The solid and dotted curves are the results of our
calculations and of the phenomenological parametrization\cite {TGA}.
The data are open circle\cite{gc27}, open square\cite{gc28},
open diamond\cite{gc29}, plus\cite {gc30}, triangle up\cite{gc31},
filled circle\cite{gc32}, and filled square\cite{gc33}, respectively.  }
\end{center}\end{figure}

\vspace{0.5cm}
\begin{figure}\begin{center}
\centering
\includegraphics[width=7.5cm, height=6cm]{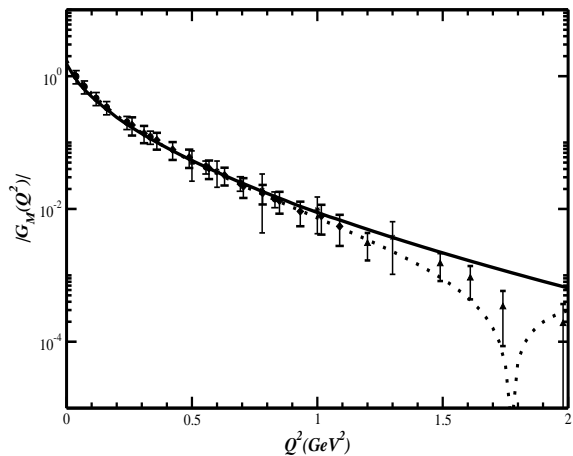}
\caption{Estimated deuteron magnetic form factor $G_{M}(Q^2)$. The solid
and dotted curves are the results of our calculations and of the
phenomenological parametrization\cite {TGA}. The
data are circle\cite{gm34}, square\cite{gm35}, diamond\cite{gm36},
and triangle\cite {gm37}, respectively.}
\end{center}\end{figure}

It should be stressed that, in this work according to the discussions of
Ref. \cite{Goldberger}, we explicitly include the $D-$ state contribution to
the deuteron-proton-neutron vertex as shown in eq. (10). Comparing to our
previous work in \cite{Dong0}, we found that this contribution is
very important for the understanding of the quadrupole moment and quadrupole
form factors. The estimated $G_M(0)$ and $G_Q(0)$ are about 1.53 and
21.38, respectively. These two values are reasonable comparing to the
normalization conditions of 1.714, and 25.83 given in eq. (4).
If we only take the $S-$ wave contribution into account, we hardly reproduce
the experimental measurement for the quadrupole moment at the zero-recoil
limit, although the estimated magnetic moment is consistent with the data.
Here, the negligibly small value of the quadrupole moment in
Ref. \cite{Dong0} is improved due to the inclusion of the $D-$ state
contribution. Meanwhile, the charge and magnetic moments also remain 
reasonably.

In summary, we explicitly consider, in this work, the $D-$ state contribution 
to the vertex of the deuteron-proton-neutron, as well as the $S-$ wave one 
simultaneously, and find that our four-dimensional phenomenological 
Lagrangian approach can reasonably reproduce the deuteron charge, magnetic, 
particularly, quadrupole form factors simultaneously. The estimated quadrupole 
moment is much improved due to the inclusion of the $D-$ state contribution. 
It should be stressed that our present
approach is a fully relativistic and it is different from the potential model
calculations based on the three-dimensional framework.

Of course, the present calculation can be further improved, since
we still cannot reproduce correctly the crossing point of the charge and
magnetic form factors of the deuteron as discussed in Ref. \cite{TGA}.
It is found that the experimental data for $G_C$ and $G_M$ show the existence
of a zero, for $Q_{0C}^2=0.7~GeV^2$ and $Q_{0M}^2=2~GeV^2$, respectively.
This is probably due to the fact that our selected correlation functions are
still simple. Moreover, the explicit form of the $D-$ state contribution, as
shown in eq. (10), is not unique \cite{Gross1}. A more sophisticated 
calculation is in progress. Finally, it is expected that the future 
calculation of the deuteron generalized parton distribution functions with 
help of this approach is promising.

\vspace{1.5cm}
\begin{figure}\begin{center}
\centering
\includegraphics[width=7.5cm, height=6cm]{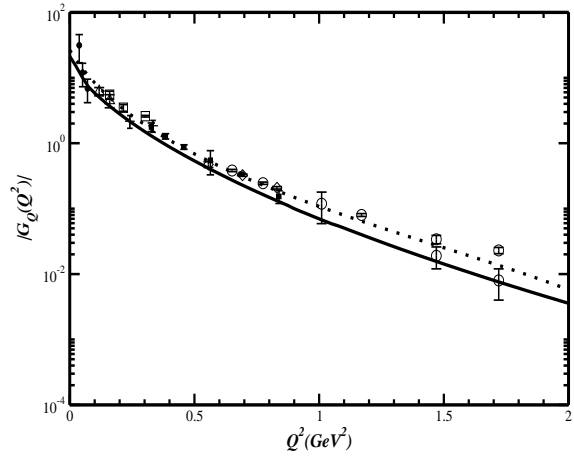}
\caption{Estimated deuteron quadrupole form factor $G_Q(Q^2)$.
Notations are the same as Fig.4. }
\end{center}\end{figure}
\vspace{0.5cm}

\begin{acknowledgments}
This work is supported  by the National Sciences Foundations
Nos. 11475192, 10975146, 11035006, and 10775148.
\end{acknowledgments}


\newpage



\end{document}